\documentclass[onecolumn,aps,preprint,showpacs]{revtex4}
\usepackage{graphicx}

\newcommand{\beq}[1]{\begin{equation}\label{#1}}
\newcommand{\beqa}[1]{\begin{eqnarray}\label{#1}}
\newcommand{\eeq}{\end{equation}}
\newcommand{\eeqa}{\end{eqnarray}}

\newcommand{\rf}[1]{(\ref{#1})}
\newcommand{\bfnabla}{\mbox{\boldmath{$\nabla$}}}

\newcommand{\bftau}{\mbox{\boldmath{$\tau$}}}

\begin{document}

\author{Yaroslav Tserkovnyak}
\affiliation{Lyman Laboratory of Physics, Harvard University, Cambridge, Massachusetts 02138}
\author{David Linton Johnson}
\affiliation{Schlumberger-Doll Research, Old Quarry Road, Ridgefield, Connecticut 06877-4108}
\author{\vspace{2cm}{\it SHORT TITLE:} {\bf Capillary forces in porous media}\\\vspace{1cm}Received\newpage}

\title{Capillary forces in the acoustics of patchy-saturated porous media\vspace{2cm}}

\begin{abstract}
A linearized theory of the acoustics of porous elastic formations,
such as rocks, saturated with two different viscous fluids is
generalized to take into account a pressure discontinuity across
the fluid boundaries. The latter can arise due to the surface
tension of the membrane separating the fluids. We show that the
frequency-dependent bulk modulus $\tilde{K}(\omega)$ for wave
lengths longer than the characteristic structural dimensions of
the fluid patches has a similar analytic behavior to the case
of a vanishing membrane stiffness and depends on the same
parameters of the fluid-distribution topology. The effect of the
capillary stiffness can be accounted for by renormalizing the
coefficients of the leading terms in the low-frequency limit
of $\tilde{K}(\omega)$.
\end{abstract}

\pacs{43.20.Jr}
\maketitle

\section{Introduction}
\label{int}

Using the Biot theory \cite{biot56} for the acoustics of
fluid-saturated porous media as a starting point, it was possible
\cite{johnson01} to formulate a simple theory for the frequency-dependent
bulk modulus, $\tilde{K}(\omega)$, of a composite consisting of a
homogeneous porous frame heterogeneously saturated with two
different viscous Newtonian fluids. It is assumed that saturation
occurs in ``patches'' of 100\% saturation of one or the other of
the fluids. In the limit when the two fluids are identical, the
theory reduces to the low-frequency regime of the Biot theory.
The mismatch of the local properties of the composite
across the patch boundaries leads to a dispersion and attenuation
mechanism \cite{johnson01} for acoustic waves, which is physically
different from that of the Biot theory. In particular, the
propagatory modes of the Biot theory, as applied to a fully
monosaturated sample, are predicted to be nondispersive and
nonattenuative in the relevant regime, $\omega\ll\omega_B$, where
$\omega_B$ is the Biot crossover frequency. In the case of patchy
saturation, on the other hand, the mismatch of the elastic
properties at the patch boundaries leads to a redistribution of
the fluids between the patches when the system is perturbed.
Because of the viscosity, this mechanism governs energy
dissipation and, therefore, the attenuation and dispersion of the
propagatory modes. There is a crossover frequency, $\omega_c$,
below which the pore pressure can equilibrate across the patches
and above which it does not.

This mechanism is particularly effective when there is a large
contrast between the two fluids as occurs when one of them is a
gas. In such a system, which was further studied in Ref.~\cite{tse02}
by analyzing measurements of Ref.~\cite{cadoret93}, the gas saturated
patches of the sample may often be approximated as vacuum.  The
relevant formulae become quite a bit simpler in this case.

While the fluid redistribution across the patch boundaries was of
a central importance in Ref.~\cite{johnson01},
its theory failed to account for the membrane surface tension which can
arise, for example, due to the pinning of the boundary edges
inside the pores. The latter will result in the formation of a
meniscus when the fluids are pushed between different patches,
which in turn leads to a pressure drop between the patches due to
the capillary forces. The present paper addresses this phenomenon.
It is important to note that various additional capillary effects,
such as the fractal nature of the fluid distribution in the pore
space, can be important in realistic porous materials but lie
outside the scope of this work.

Johnson \cite{johnson01} suggested that the patchy-saturation effect may
be a dominant mechanism for the low-frequency
($\omega\ll\omega_B$) attenuation and dispersion in carbonate
rocks, but can be dwarfed by microscopic ``squirt'' mechanisms in
materials like sandstones. We indeed
later showed \cite{tse02} that the sonic and even ultrasonic measurements
\cite{cadoret93,cadoret95,cadoret98} on various limestones can be
successfully described by the patchy-saturation effect. In
particular, we demonstrated there how one can extract information
about the characteristic patch sizes and shapes in partially water
saturated limestones by measuring the velocity and attenuation of
acoustic waves. Extending our theory to take into account the
capillary forces is necessary to improve our understanding of the
acoustics in patchy-saturated materials. While the effect of the
membrane tension at the boundary is but one of the important
effects, investigating its effects can be a starting point for
further studies of the capillary effects in porous media.

The manuscript is structured as follows. In Sec.~\ref{gen} we formulate the assumptions and limitations of the theory. Then, in Sec.~\ref{fa} we in turn discuss the static, low-frequency, and high-frequency regimes of the theory in the general case of the patchy saturation by two arbitrary fluids. In Sec.~\ref{fv} an alternative formulation of the static and low-frequency limits is presented which is simple and physically transparent, albeit only valid for the case when one of the fluids is a gas. We compare our findings with exact numerical calculations in Sec.~\ref{ex} for two simple geometries: that of the concentric spheres and the periodic slab geometries. Finally, in Sec.~\ref{tm} a simple theoretical model for $\tilde{K}(\omega)$ is presented enabling us to connect the low- and high-frequency asymptotes of the theory. Our results are then summarized in Sec.~\ref{con}. Two somewhat technical discussions, on the capillary correction to the high-frequency limit and analytic structure of $\tilde{K}(\omega)$, are given separately from the main text, in Appendices~\ref{a1} and \ref{a2}, respectively.

\section{Generalities}
\label{gen}

As in Ref.~\cite{johnson01}, we consider a porous rock fully saturated
with two different fluids, one having global saturation $S_1$ and the
other $S_2=1-S_1$. Each point of the sample is saturated with
one fluid or the other and the local elastic properties of the
composite are assumed to be governed by the usual Biot parameters
of the rock skeleton and the locally present fluid. The only
exception is the patch and the sample boundaries which deserve
special treatment (in the following).
While we will talk about various limits of the
frequency dependence of the bulk modulus $\tilde{K}(\omega)$, the
Biot theory is always assumed to be in its low-frequency regime.
The Biot crossover frequency \cite{biot56} \beq{ob}
\omega_B=\frac{\eta\phi}{k\rho_f\alpha_\infty} \eeq
(using the notation of Ref.~\cite{johnson94}), defined in
terms of the fluid density $\rho_f$, the viscosity $\eta$, the
tortuosity of the pore space $\alpha_\infty$, the dc permeability
$k$, and the porosity $\phi$, is thus the upper frequency limit of
our theory, i.e., it is assumed that $\omega\ll\omega_B$. Biot
equations of motion within each fluid patch \cite{biot56} then
simplify to \beqa{m1}
\bfnabla\cdot\bftau&=&0\,,\\
\bfnabla
p_f&=&\frac{i\omega\phi\eta}{k}(\mathbf{U}-\mathbf{u})\,,\label{m2}
\eeqa where all quantities vary as $\mathbf{u}(\mathbf{
r})e^{-i\omega t}$. Here, $p_f$ is the acoustic component of the pore
pressure, $\mathbf{U}$ and
$\mathbf{u}$ are the displacement of the fluid and solid phases, respectively,
and $\bftau$ stands for the total stress due to both (solid and fluid)
phases. Note that Eqs.~\rf{m1}, \rf{m2} do not contain the inertia
terms, which is a relevant approximation as the
attenuation/dispersion mechanism of our theory is governed by the
slow diffusive mode.
The pore pressure $p_f$ and the total stress $\bftau$ are
related to the deformation strain in the solid phase,
$\epsilon_{ij}=(1/2)[u_{i,j} + u_{j,i}]$, and to that in the fluid
phase, $E_{ll}=U_{l,l}$, by \cite{biot56} \beqa{bio}
\bftau&=&\left[(P+Q-2N)\epsilon_{ll}+(R+Q)E_{ll}\right]\delta_{ij}+2N\epsilon_{ij}\,,\\
p_f&=&-\frac{1}{\phi}[Q\epsilon_{ll}+RE_{ll}]\label{bio2}\,, \eeqa
where the coefficients $P$, $Q$, and $R$ depend on the bulk moduli
of the solid phase, $K_s$, the porous frame, $K_b$, and the
pore fluid, $K_f$, as well as the porosity and the shear modulus
of the porous skeleton, $N$ (see, e.g, Ref.~\cite{johnson86}). The fast
compressional and shear velocities of the Biot theory are
nondispersive and nonattenuative whereas the slow compressional
wave is diffusive in the low-frequency limit considered here.
Furthermore, we assume that the wave lengths of the fast and shear
waves are larger than the characteristic fluid-patch dimension
$L$, i.e., $\omega\ll\omega_x$, where \beq{ox} \omega_x=\frac{2\pi
V_{\rm sh}}{L} \eeq with $V_{\rm sh}$ being the (slower)
shear-wave velocity. The latter assumption will allow us to define
the effective bulk modulus $\tilde{K}(\omega)\stackrel{\rm
def}{=}-V(P_e/\delta V)$, where $(\delta V/V) e^{-i\omega t}$ is
the oscillatory fractional volume change in response to an
external normal stress, $\bftau\cdot\mathbf{\hat{n}}=-P_e\mathbf
{\hat{n}}e^{-i\omega t}$, $\mathbf{\hat{n}}$ being the outward
normal of the surface. Finally, the sample is considered to be
isotropic and homogeneous, except for the fluid patches.

Due to capillary forces, the fluid pore pressure is in general
discontinuous across patch boundaries.  Absent any applied stress,
the pressure discontinuity across the boundary, i.e., the capillary
pressure,  $P_f^{(1)}-P_f^{(2)}$, is related to the mean curvature of
the interface within each pore and to the surface tension via the
Laplace-Young equation.  In a completely different context than we
are considering in this article, one is led to the equations of
two-phase flow in which the partial saturation is generally
treated as a well-defined function of the capillary pressure. See,
e.g., Bear~\cite{bear}. These concepts make no distinction between partial
saturations that are patchy and those in which each pore may be
partially saturated.  Indeed, $P_f^{(1)}$ and $P_f^{(2)}$ are each
generally considered to be macroscopic quantities defined at each
material point throughout the system.

In the present context, however, we assume the fluid patches are
perturbed from equilibrium by the application of an additional
small-amplitude acoustic wave.  The changes in the pore pressures
are small compared to the equilibrium values: $|p_f({\bf r})|
\ll|P_f^{(j)}|,~\{j= 1,2\}$, where $p_f({\bf r})$ is
the acoustic component of the pore pressure.  On a microscopic
level, the meniscus bends in and out, relative to its rest
position.  On a macroscopic level, there is an oscillatory
displacement of fluid relative to the solid frame, at the boundary
of the patch. This effect is conventionally captured by
postulating the following boundary condition at the interface
between two patches: \beq{dpf}
p_f^{(1)}-p_f^{(2)}=W\phi(\mathbf{U}-\mathbf{u})\cdot\mathbf{\hat{n}}\,. \eeq
Here, $p_f^{(j)}$ is the acoustic pore pressure on the $j$th side of
the boundary surface and $\mathbf{\hat{n}}$ is an outward normal
from patch 1 to patch 2. $\mathbf{U}\cdot\mathbf{\hat{n}}$ and
$\mathbf{u}\cdot\mathbf{\hat{n}}$ represent the components of the fluid
and the solid displacements, respectively, that are normal to the
patch surface; each of these entities is separately continuous
across the patch boundary. Thus, Eq.~\rf{dpf} simply states that
the additional amount by which the fluid interface bulges is
proportional to the additional capillary pressure induced by the
acoustic wave; the spring constant, $W$, is often called the
``membrane stiffness".  Indeed, Nagy and Blaho~\cite{nagy94} have shown
that Eq. \rf{dpf} follows from the Laplace-Young equation applied
to the case of pores that are circular cylinders. In that example
they showed that $W = \sigma/k$, where $\sigma$ is
the surface tension between the two fluids.  This specific example
illustrates the point that surface tension manifests itself as a
stiffer membrane in smaller pores than in larger ones. More generally,
the membrane stiffness is predicted to take the form of
\beq{YT4} W = s\sigma/k\,, \eeq where $s$ is a shape factor which
becomes much less than one as the pore structure becomes more
irregular \cite{nagy94}.

Our previous results (Johnson~\cite{johnson01},
Tserkovnyak and Johnson~\cite{tse02}) were specific to the case $W\equiv 0$.
Here, we treat $W$ as a real-valued, frequency-independent, phenomenological
parameter of the theory, to be measured independently in a given
context. Nagy and Blaho~\cite{nagy94} have done just that for various
porous media. See also Ref.~\cite{nagy95}. Furthermore, Nagy~\cite{nagy92}
successfully used Eq.~\rf{dpf} to analyze his data on surface
acoustic modes of fully-saturated porous media.  In a somewhat
different context, Eq.~\rf{dpf} is used by Liu and Johnson~\cite{liu97}
to describe the effect of an elastic mudcake on tube waves in
permeable formations.  The approximate validity of Eq.~\rf{dpf} is
established on firm theoretical and experimental grounds.  It is
the essential purpose of the present article to investigate the
effects of a nonzero value for $W$ on the acoustic properties of
patchy-saturated samples.

Eq.~\rf{dpf} can be combined with
Eq.~\rf{m2} to generalize the latter to hold at the interfaces as
well as inside the patches: \beq{m2g} \bfnabla p_f+W\phi[\mathbf{
\hat{n}}\cdot(\mathbf{U}-\mathbf{u})]\mathbf{
\hat{n}}\delta(R)=\frac{i\omega\phi\eta}{k}(\mathbf{U}-\mathbf{
u})\,, \eeq where $R(\mathbf{r})$ is the shortest distance from a
given point $\mathbf r$ to the patch boundary and $\delta(R)$ is
the Dirac's $\delta$ function. In addition, we assume the no net
flow condition over the surface of the sample: \beq{sb} \int
dS[\mathbf{U}-\mathbf{u}]\cdot\mathbf{\hat{n}}=0\,, \eeq which
is valid either for sealed outer boundary or for the periodic
boundary conditions. (The integration is taken either over the
surface of the sample, for the sealed-pore boundary condition, or
over a unit cell, for the periodic boundary conditions.)
Eq.~\rf{sb} can be converted to a volume integral using Gauss'
theorem: \beq{Y3b} \langle\epsilon_{ll}\rangle=\langle
E_{ll}\rangle\,. \eeq

To summarize this section, the present theory supercedes that of
Ref.~\cite{johnson01} in that the assumption of vanishing capillary
forces is now relaxed. The capillarity is accounted for by considering
the pressure drop across interfaces which is governed by membrane
stiffness, see Eq.~\rf{dpf}. The theory of Ref.~\cite{johnson01} is
thus recovered in the limit when $W\rightarrow0$.

\section{High and Low Frequency Limits}
\label{fa}

\subsection{Static regime}

In the static limit, it is possible to get an exact closed-form
analytic expression for the compressive modulus, $K_0$, defined by
Eqs.~\rf{m1}-\rf{dpf}. Hill \cite{hill63} has developed a well-known
exact result for the bulk modulus of a (nonporous) composite in
which the shear modulus is everywhere constant, though the bulk
modulus varies from point to point. See also Ref.~\cite{milton}. We
extend that theory to the present case by looking for solutions of
the form \beq{Y5} \left.\begin{array}{rcl}
u_i&=&\alpha x_i+\chi_{,i}\,,\\
U_i&=&\alpha x_i+\psi_{,i}\,.
\end{array}\right.
\eeq
It is understood that $\chi$ as well as the normal displacement
$\chi_{,i}n_i$, are continuous across the patch boundary;
similarly for $\psi$ and $\psi_{,i}n_i$. Inasmuch as the first
terms in Eqs.~\rf{Y5} are special cases of the second terms, one
is free to choose any value for $\alpha$, including zero.
It is slightly convenient, however, to choose
\beq{Y6}
3\alpha=\langle\epsilon_{ll}\rangle\equiv\langle E_{ll}\rangle\,.
\eeq
Eq.~\rf{bio} can now be written, within any patch, as
\beq{Y7}
\tau_{ij,j}=\left[(P+Q)\chi_{,jj}+(R+Q)\psi_{,jj}\right]_{,i}\stackrel{\rm set}{=}0\,.
\eeq
This equation is satisfied if the Laplacians of $\chi$ and of $\psi$
are constant within each patch:
\beqa{Y8}
\chi_{,jj}&=&\left\{\begin{array}{ll}
a_1 & {\bf r}\in{\rm patch\,1}\\
a_2 & {\bf r}\in{\rm patch\,2}
\end{array}\right.\,,\\
\psi_{,jj}&=&\left\{\begin{array}{ll}
b_1 & {\bf r}\in{\rm patch\,1}\\
b_2 & {\bf r}\in{\rm patch\,2}
\end{array}\right.\,.
\eeqa The four constants, $\{a_1, a_2, b_1, b_2 \}$, are
determined by various ancillary conditions. First, the volumetric
strains within the $j$th patch may be written as \beq{Y9}
\begin{array}{rcl}
\epsilon^{(j)}_{ll}&=&3\alpha+a_j\,,\\
E^{(j)}_{ll}&=&3\alpha+b_j\,.
\end{array}
\eeq
Therefore, Eq.~\rf{Y6} gives two conditions on
the constants:
\beq{Y10}
\begin{array}{rcl}
S_1 a_1+S_2 a_2&=&0\,,\\
S_1 b_1+S_2 b_2&=&0\,.
\end{array}
\eeq

Next, there is the condition that the traction,
$\bftau\cdot\mathbf{\hat{n}}$, is continuous across the boundary
between patches. Hill \cite{hill63} has pointed out an identity
satisfied by the discontinuity across the patch boundary of the
second derivatives of $\chi$ which can be rewritten for our
purposes as \beq{Y12}
\left[\chi_{,ij}^{(1)}-\chi_{,ij}^{(2)}\right]n_j=(a_1-a_2)n_i\,,
\eeq where the superscript on $\chi^{(k)}$ refers to the $k$th
patch. Therefore, the discontinuity in the traction is \beqa{Y13}
\left[\tau_{ij}^{(1)}-\tau_{ij}^{(2)}\right]n_j&=&\left\{\left[(P_1+Q_1-2N)(3\alpha+a_1)+(R_1+Q_1)(3\alpha+b_1)\right]\right.\nonumber\\
&&-\left[(P_2+Q_2-2N)(3\alpha+a_2)+(R_2+Q_2)(3\alpha+b_2)\right]\nonumber\\
&&+\left.2N(a_1-a_2)\right\}n_i\,.
\eeqa
($P_k, Q_k, R_k$ refer to $P, Q, R$ evaluated with respect
to the saturating fluid in patch $k$.)
In order for Eq.~\rf{Y13} to vanish across the patch
interface, it is necessary that the quantity in curly brackets vanishes:
\beq{Y14}
(P_1+Q_1)(3\alpha+a_1) +(R_1+Q_1)(3\alpha+b_1)=(P_2+Q_2)(3\alpha+a_2)+(R_2+Q_2)(3\alpha+b_2)\,.
\eeq
This, then, is the third condition on $\{a_1, a_2, b_1, b_2 \}$.

The fourth and final condition follows from Eq.~\rf{dpf}.
The pore pressure within the $k$th patch is given by
Eqs.~\rf{bio2} and \rf{Y9}:
\beq{Y15}
p_f^{(k)}=-\frac{1}{\phi}[R_k(3\alpha+b_k)+Q_k(3\alpha+a_k)]\,.
\eeq
We see that the pore pressure is constant within each patch, as required
by Eq.~\rf{m2} in the static limit. The left-hand side of the boundary
condition \rf{dpf} is constant over the interface
between the two patches. Consider the case of a finite volume with
sealed boundaries. Integrating Eq.~\rf{dpf} over the surface
between the two patches, one finds
\beq{Y16}
\left[p_f^{(1)}-p_f^{(2)}\right]A=W\phi\int_{A}dA(\mathbf{U-u})\cdot\mathbf{\hat{n}}\,,
\eeq
where $A$ is the area of the interface between the patches.
This integration can be extended over the entire bounding surface of
fluid 1, say, because there is no flow across the outer boundary:
\beq{Y17}
\left[p_f^{(1)}-p_f^{(2)}\right]A=W\phi\int_{A_1}dA(\mathbf{U-u})\cdot\mathbf{\hat{n}}\,.
\eeq
Now one may use Gauss' theorem:
\beq{Y18}
\left[p_f^{(1)}-p_f^{(2)}\right]A=W\phi\int_{V_1}dV(E_{ll}-\epsilon_{ll})=W\phi S_1V\left[E_{ll}^{(1)}-\epsilon_{ll}^{(1)}\right]\,.
\eeq
The volumetric strains and the pore pressure can be related to the unknown
constants by means of Eqs.~\rf{Y9} and \rf{Y15}, respectively.
The result for Eq.~\rf{Y18} is then
\beq{Y19}
R_1(3\alpha+b_1)+Q_1(3\alpha+a_1)-\left[R_2(3\alpha+b_2)+Q_2(3\alpha+a_2)\right]=\tilde{W}(a_1-b_1)S_1\,,
\eeq
where
\beq{Y19b}
\tilde{W}=W\phi^2\frac{V}{A}
\eeq
in terms of the sample volume, $V$, and the bounding area between the patches,
$A$. If the derivation is repeated for the case of periodic boundary
conditions, Eq.~\rf{Y19} still holds; in that case, $V$ is the
volume of a unit cell and $A$ is the interface area between the
patches within the unit cell. In either case, if the derivation
is repeated by closing the surface over fluid 2, instead of fluid 1,
the right-hand side of Eq.~\rf{Y19} is replaced by $-\tilde{W}(a_2-b_2)S_2$,
which is equivalent via Eqs.~\rf{Y10}.

The four linear equations, \rf{Y10}, \rf{Y14}, and \rf{Y19}, can
be inverted analytically to get the ratios $\{a_1, a_2, b_1,
b_2\}/\alpha$; in practice, this is more easily done numerically.
The effective static modulus is simply related to these constants.
Following Ref.~\cite{hill63}, we identify the external pressure, $P_e$,
with the average compressive stress in the system: \beq{Y20}
P_e=-\frac{1}{3}\langle\tau_{ll}\rangle=-\frac{1}{3}\left[S_1\tau_{ll}^{(1)}+S_2\tau_{ll}^{(2)}\right]\,,
\eeq where the compressive stress in the $k$th patch follows from
Eq.~\rf{bio}: \beq{Y21}
\tau_{ll}^{(k)}=3\left[\left(P_k+Q_k-\frac{4}{3}N\right)a_k+(R_k+Q_k)b_k+3K^{(k)}_{\rm
BG}\alpha\right]\,. \eeq Here, $K_{\rm
BG}^{(k)}=P_k+2Q_k+R_k-(4/3)N$ is the Biot-Gassmann modulus for
the $k$th patch. The effective static modulus for the system,
$K_0=-P_e/\langle\epsilon_{ll}\rangle$, is thus given by
\beqa{Y22}
K_0&=&\frac{1}{3}\left\{S_1\left[\left(P_1+Q_1-\frac{4}{3}N\right)\frac{a_1}{\alpha}+(R_1+Q_1)\frac{b_1}{\alpha}+3K_{\rm BG}^{(1)}\right]\right.\nonumber\\
&&+\left.S_2\left[\left(P_2+Q_2-\frac{4}{3}N\right)\frac{a_2}{\alpha}+(R_2+Q_2)\frac{b_2}{\alpha}+3K_{\rm BG}^{(2)}\right]\right\}\,.
\eeqa

Within the context of the model, Eq.~\rf{Y22} is exact. It depends
upon the usual Biot parameters, the saturation values, the
membrane stiffness, $W$, and it also depends upon the surface to
volume ratio of the patch interface, $A/V$. There are three
interesting limiting cases of Eq.~\rf{Y22}: (1) If the two fluids
have identical bulk moduli, then \beq{Y23}
\lim_{K_f^{(1)}=K_f^{(2)}}K_0=K_{\rm BG}^{(1)}=K_{\rm
BG}^{(2)}\,, \eeq independent of the value of the membrane
stiffness parameter, $W$. When the sample is compressed, the pore
pressure rises equally in the two patches with the result that
there is no relative flow between fluid and solid. (2) If the
membrane stiffness is large enough, then each fluid is locked
within its own patch and the static modulus is equal to the
Biot-Gassmann-Hill result: \beq{Y24} \lim_{W\rightarrow
\infty}K_0=K_{\rm BGH}\,, \eeq where \beq{bgh} \frac{1}{K_{\rm
BGH}+(4/3)N}=\frac{S_1}{K_{\rm BG}^{(1)}+(4/3)N}+\frac{S_2}{K_{\rm
BG}^{(2)}+(4/3)N}\,. \eeq (3) If the membrane stiffness is zero,
then the static modulus is given by the Biot-Gassmann-Woods result
as was proved previously \cite{johnson01}: \beq{Y25}
\lim_{W\rightarrow0}K_0=K_{\rm BGW}\,, \eeq where $K_{\rm
BGW}=K_{\rm BG}(K_f^{\rm eff})$ and $K_f^{\rm eff}$ is an
effective fluid modulus given by the Wood's law expression:
\beq{kw} \frac{1}{K_f^{\rm
eff}}=\frac{S_1}{K_f^{(1)}}+\frac{S_2}{K_f^{(2)}} \:\:\:. \eeq

In Fig.~\ref{fig:Y1} we show the behavior of $K_0$ as a function
of $\tilde{W}$. The parameter set \cite{table} is based on that of
Ref.~\cite{johnson01}. The sample is
saturated with water, $S_w=0.7$, and with air, $S_a=0.3$. If the
membrane stiffness is small, then Eq.~\rf{Y25} holds whilst if it
is large, Eq.~\rf{Y24} is valid. The crossover occurs when
$\tilde{W}$ is approximately equal to the modulus of the stiffer
fluid, water. In Fig.~\ref{fig:Y2} we show the saturation
dependence of the static modulus, $K_0$, as well as that of
$K_{\rm BGW}$ and $K_{\rm BGH}$. Even when the membrane-stiffness
parameter, $W$, is independent of the saturation, the
parameter $\tilde{W}$ will in general have a saturation dependence
reflecting the change in the interfacial area. Nonetheless, for
this plot, $\tilde{W}$ is held constant at a value
$\tilde{W}=1$~GPa.

\subsection{Low-frequency limit}

Here we wish to find the behavior of $\tilde{K}(\omega)$ as it
approaches $K_0$. Using the method developed in Ref.~\cite{johnson01},
we consider two equivalent expressions for the average energy
dissipation per cycle. The macroscopic expression for the
dissipation power averaged over one cycle is \beq{Y26}
\bar{P}=\frac{1}{2}{\rm Real}\int dS\frac{\partial\mathbf{
u^\ast}}{\partial
t}\cdot\bftau\cdot\mathbf{\hat{n}}=-\frac{1}{2}{\rm
Real}\left[i\omega V\frac{|P_e|^2}{\tilde{K}(\omega)}\right]\,,
\eeq where the integral is taken over the bounding surface of the
sample. The microscopic expression is \beq{Y27}
\bar{P}=-\frac{1}{2}{\rm Real}\int
dV\phi\left(\frac{\partial\mathbf{U^\ast}}{\partial
t}-\frac{\partial\mathbf{u^\ast}}{\partial t}\right)\cdot\bfnabla
p_f\,. \eeq In Eq.~\rf{Y27} the integration is taken over the
volumes of each patch, specifically excluding the bounding
surfaces of the patches; $\bfnabla p_f$ has a $\delta$-function
contribution on the patch surface, see Eq.~\rf{m2g}. One may then
substitute Eq.~\rf{m2} into Eq.~\rf{Y27} to arrive at the
following microscopic expression for the dissipation: \beqa{Y29}
\bar{P}&=&-\frac{1}{2}{\rm Real}\int dV\phi i\omega(\mathbf{U^\ast}-\mathbf{u^\ast})\cdot\left[\frac{i\omega\phi\eta}{k}(\mathbf{U}-\mathbf{u})\right]\nonumber\\
&=&\frac{\omega^2\phi^2}{2k}\int dV\eta(\mathbf{r})\left|\mathbf{U}(\mathbf{r})-\mathbf{u}(\mathbf{r})\right|^2\,.
\eeqa
In this expression, the integrand does not contain any $\delta$-function
contributions. It is immediately clear that, as expected,
the low-frequency behavior of $\bar{P}$ is quadratic in frequency.
Therefore, the low-frequency expansion of $\tilde{K}(\omega)$ is given by
\beq{Y30}
\lim_{\omega\rightarrow0}\tilde{K}(\omega)=K_0\left[1-i\omega T+O(i\omega)^2\right]\,,
\eeq
where the parameter $T$ is determined by substituting Eq.~\rf{Y30}
into Eq.~\rf{Y26} and equating it to Eq.~\rf{Y29}:
\beq{Y31}
T=\frac{\phi^2K_0}{kV|P_e|^2}\int dV\eta(\mathbf{r})\left|\mathbf{U}^{(0)}(\mathbf{r})-\mathbf{u}^{(0)}(\mathbf{r})\right|^2\,,
\eeq
which is expressed in terms of the static displacements,
$\mathbf{U}^{(0)}$ and $\mathbf{u}^{(0)}$.

We know from Eqs.~\rf{Y5} that the displacements in the static
limit are given by the gradient of certain functions. Therefore,
we may define an auxiliary function, $\Phi$, by \beq{Y32}
\bfnabla\Phi=-\eta(\mathbf{r})\frac{\mathbf{U}^{(0)}-\mathbf{
u}^{(0)}}{P_e}\,. \eeq Within each patch $\Phi = \eta [\chi -
\psi]/P_e + {\rm const}$. The function $\Phi$ is continuous across
a patch boundary (otherwise there would be a $\delta$-function
contribution to the displacements); the normal component of the
vector $-\bfnabla\Phi/\eta(\mathbf{r})$ is also continuous;
and $\Phi$ is either periodic or
$\bfnabla\Phi=0$ (in the case of the sealed-pore boundary condition)
on the outer boundary. The
function $\Phi$ satisfies the differential equation \beq{Y33}
\bfnabla\cdot\left\{\frac{-1}{\eta(\mathbf{
r})}\bfnabla\Phi\right\}=\frac{E_{ll}^{(0)}(\mathbf{
r})-\epsilon_{ll}^{(0)}(\mathbf{r})}{P_e}\stackrel{\rm
def}{=}g(\mathbf{r})\,. \eeq This equation is analogous to an
electrical conductivity problem in which $\Phi(\mathbf{r})$ plays
the role of a potential, $\eta(\mathbf{r})$ is the local
resistivity, and $g(\mathbf{r})$ is the distributed current
source. Electrical neutrality is guaranteed by Eq.~\rf{Y3b}: $\int
dVg(\mathbf{r})=0$.

In terms of $\Phi$, Eq.~\rf{Y31} may be written as \beq{Y33a}
T=\frac{\phi^2K_0}{kV}\int dV\frac{1}{\eta(\mathbf{
r})}(\bfnabla\Phi)^2\,, \eeq which may be integrated by parts:
\beq{Y34} T=\frac{\phi^2K_0}{kV}\int dVg(\mathbf{r})\Phi(\mathbf{
r})\,. \eeq (The surface term vanishes identically for either the
sealed or the periodic boundary condition.) The source function,
$g(\mathbf{r})$, is constant within each patch but in general is
discontinuous across the patch interfaces. From Eqs.~\rf{Y5},
\rf{Y8}, and \rf{Y9} one has \beq{Y35}
g(\mathbf{r})=\left\{\begin{array}{ll}(b_1-a_1)/P_e&\mathbf{r}\in{\rm patch\,1}\\
(b_2-a_2)/P_e&\mathbf{r}\in{\rm patch\,2}
\end{array}\right.\,.
\eeq
Thus, the determination of the parameter $T$ which
governs the low-frequency approach to the static limit, Eq.~\rf{Y30},
requires solving the differential equation \rf{Y33}
in which the piecewise constant sources, Eq.~\rf{Y35},
are known by virtue of the solution for the static regime,
and imposing one of our two no net-flow boundary conditions. This situation is
only slightly different from that encountered when the surface-membrane
effect is ignored altogether \cite{johnson01}.

\subsection{High-frequency limit}

Here we consider frequencies that are high in the sense that the
pore pressure does not have time to equilibrate from patch to
patch but we are still in the regime where $\omega \ll \omega_B$
and $\omega \ll \omega_x$. In this high frequency limit,
$\omega\rightarrow\infty$, the fluids within each patch are locked
with the solid frame, $\epsilon_{ll}=E_{ll}$. This conclusion is
independent of the interface-stiffness parameter, $W$.  We have a
situation which satisfies the assumptions of Ref.~\cite{hill63}: The
shear modulus is a global constant and the bulk modulus is
constant within each patch. Hence, \beq{if}
K_\infty\stackrel{\rm def}{=}\lim_{\omega\rightarrow\infty}\tilde{K}(\omega)=K_{\rm
BGH} \eeq for any value of $W$, where $K_{\rm BGH}$ is given by
Eq. \rf{bgh}. Recall that we found the same result for the static
compressive modulus, $K_0$, when the patch boundary is locked by
the large capillary tension, $W\rightarrow\infty$, see
Eq.~\rf{Y24}.

For a large but finite frequency, the correction to $K_\infty$ is
provided by the diffusive modes at the patch boundary
\cite{johnson01}. In terms of a local coordinate $x$ normal to the
interface, the pore pressure near the boundary, $x=0$, is given by
\beq{sw}
p_f=\left\{\begin{array}{lr}p_{f1} +A_1e^{-iq_1x},&x<0~{\rm~(patch\,1)}\\
p_{f2} + A_2e^{iq_2x},&x>0~{\rm~(patch\,2)}\end{array}\right.\,.
\eeq Here, $p_{fk}$ is the pore pressure inside the $k$th patch in
the limit of infinite frequency and \beq{swv}
q_k=\sqrt{i\omega/D_k} \eeq is the slow-mode wave vector in terms
of the diffusion coefficient $D_k$; see Appendix~\ref{a1}. In the
case of a finite membrane stiffness, $p_f(\mathbf{r})$ is in
general discontinuous across the patch boundary. We can,
nevertheless, show (see Appendix~\ref{a1}) that the pore pressure
discontinuity $\delta p_f \stackrel {\rm def}{=}|p_f^{(1)}-p_f^{(2)}|=
O(\omega^{-1/2})\ll \Delta p_f \stackrel {\rm def}{=}
|p_{f1}-p_{f2}|=O(\omega^0)$, for large frequencies, and thus the
leading correction to $K_\infty$ can be found by setting $\delta
p_f=0$, i.e., assuming continuity of the pressure, Eq.~\rf{sw}.
Requiring, in addition, continuity of the fluid flux, we can then
find coefficients $A_{1,2}$ in Eq.~\rf{sw} and reproduce the
result of Ref.~\cite{johnson01}: \beq{hf}
\lim_{\omega\rightarrow\infty}\tilde{K}(\omega)=K_{\rm
BGH}\left[1-(i\omega)^{-1/2}G+\cdots\right]\,, \eeq where \beq{g}
G=\frac{kK_{\rm
BGH}}{\eta_1\sqrt{D_1}+\eta_2\sqrt{D_2}}\left(\frac{\Delta
p_f}{P_e}\right)^2\frac{A}{V} \eeq in terms of the would-be drop
in the pore pressure, $\Delta p_f$, relative to the applied
external stress, $P_e$: \beq{dp} \frac{\Delta
p_f}{P_e}=\frac{(R_2+Q_2)[K_1+(4/3)N]-(R_1+Q_1)[K_2+(4/3)N]}{\phi
S_1K_1[K_2+(4/3)N]+\phi S_2K_2[K_1+(4/3)N]}\,. \eeq Notice that
the membrane-stiffness parameter does not enter any of
Eqs.~(\ref{hf})-(\ref{dp}), so that the capillary forces not only
have vanishing contribution to the high-frequency compressive
modulus \rf{if} but also to the leading correction to it,
Eq.~\rf{hf}.

\section{Liquid and vacuum}
\label{fv}

In Sec.~\ref{fa} we have seen that whereas the two terms
in the high frequency limit, Eqs.~\rf{hf}, are identical to those
derived earlier when $W=0$, the two terms which describe
the low frequency limit, Eqs.~\rf{Y30}, are now considerably more
complicated. The prescription for finding the static bulk
modulus, $K_0$ [Eq.~\rf{Y22}], and the coefficient of the leading
low-frequency correction, $T$ [Eq.~\rf{Y34}], follows from the
solution of the system of equations for parameters
$\{a_1,a_2,b_1,b_2\}$; no simple analytic form is deduced and no
transparent physical interpretation is readily available. We
therefore offer an alternative derivation, which is both simple
and physically appealing, albeit valid only for the case when one
of the two fluids is a gas taken here to be the vacuum, and the
other is a liquid whose properties are similar to water or oil. (A
gas at standard temperature and pressure
is four orders of magnitude more compressible than
water or oil and so the approximation is a good one except for gas
saturations less than $\sim10^{-4}$.)

As in Sec.~\ref{fa}, the starting point is Hill's theory
\cite{hill63}: If we denote the bulk modulus of the patch
saturated with the liquid by $K^\prime$, then the static modulus
of the composite is given by \beq{K0}
\frac{1}{K_0+(4/3)N}=\frac{S}{K^\prime+(4/3)N}+\frac{1-S}{K_b+(4/3)N}\,,
\eeq where $S$ is the fluid saturation and $K_b$ is the bulk
modulus of the dry frame, as before. $K^\prime$ can be found by
extending the Biot theory: Since the liquid does not bear any
shear stresses, the static compressive modulus of the
liquid-saturated patch must be given by the Biot-Gassmann formula,
\beq{kvbgh} K^\prime=K_{\rm BG}(K_f^\prime) \:\:\:,\eeq for some
effective fluid bulk modulus, $K_f^\prime$. This effective modulus
characterizes the ``effort of squeezing'' the liquid out of the
pore space. In particular, two limiting cases are obvious. (1)
$K_f^\prime$ vanishes with the membrane stiffness: \beq{kw0}
\lim_{W\rightarrow0}K_f^\prime=0 \eeq and (2) it reduces to the
bare fluid modulus for infinitely large stiffness parameter:
\beq{kwi} \lim_{W\rightarrow\infty}K_f^\prime=K_f\,. \eeq If by
inducing the pore pressure $p_f$, the fluid-filled pore-space
volume is reduced by $\delta V$, then $K_f^\prime$ is defined by
\beq{kf1} \frac{1}{K_f^\prime}=\frac{1}{p_f}\frac{\delta V}{\phi
SV}\,. \eeq If the fluid volume squeezed out of the pore space
into menisci formed at the patch surface is $\delta V^\prime$,
then \beq{kf2} p_fA=W\delta V^\prime\,, \eeq which follows from
Eq.~\rf{Y17}. Finally, we have \beq{kf3}
\frac{1}{K_f}=\frac{1}{p_f}\frac{\delta V-\delta V^\prime}{\phi
SV}\,. \eeq for the bare bulk modulus of the pore fluid. Combining
Eqs.~\rf{kf1}-\rf{kf3} we then arrive at \beq{kfp}
\frac{1}{K_f^\prime}=\frac{1}{K_f}+\frac{A}{W\phi SV}\,, \eeq
which satisfies Eqs.~\rf{kw0} and \rf{kwi} in the two limits given
by Eqs.~\rf{kw0} and \rf{kwi} . As a consequence, in the limit
$W\rightarrow0$, $K^\prime=K_b$ (which is a specific case of a
general result holding for arbitrary two fluids: $K^\prime=K_{\rm
BGW}$), and for $W\rightarrow\infty$, $K^\prime=K_{\rm BGH}$.
Equations~\rf{K0}, \rf{kvbgh}, and ~\rf{kfp} complete the
derivation for the static limit. This result for the static
modulus is also plotted in Figs.~\ref{fig:Y1} and \ref{fig:Y2}
where it is seen to be identical to that computed from the full
formalism.

It is as straightforward to find the parameter $T$ governing the
low-frequency behavior. $T$ is given by Eq.~\rf{Y34} after solving
Eq.~\rf{Y33}, when the function $g(\mathbf{r})$ is known. In the
fluid-filled patch, the volumetric strain of the solid can be
found using Hill's theory once again (see also Eq.~(35) of
Ref.~\cite{johnson01}):
\beq{vs1} \epsilon_{ll}=-\frac{[K_b+(4/3)N]P_e}{K^\prime
K_b+(4/3)N[SK^\prime+(1-S)K_b]}\,. \eeq Using Eq.~\rf{bio2} and
\beq{vs2} p_fA=W\phi SV(E_{ll}-\epsilon_{ll})\,, \eeq provided by
Eq.~\rf{Y18}, we then obtain \beq{vs3}
E_{ll}-\epsilon_{ll}=-\frac{Q+R}{\tilde{W}S+R}\epsilon_{ll}\,,
\eeq which together with Eq.~\rf{vs1} defines the value of $g$,
Eq.~\rf{Y33}, inside the fluid patch: \beq{gv}
g=\left(\frac{Q+R}{\tilde{W}S+R}\right)\frac{[K_b+(4/3)N]}{K^\prime
K_b+(4/3)N[SK^\prime+(1-S)K_b]}\,. \eeq Here, quantities $Q$ and
$R$ are evaluated using the bare fluid bulk modulus, $K_f$.
Defining an auxiliary function $\tilde{\Phi}$ in the space
occupied by the fluid patch, $V_f$, by \beq{J23}
\nabla^2\tilde{\Phi}=-1~~~\mathbf{r}\in V_f \eeq subject to the
patch boundary condition $\tilde{\Phi}=0$
as well as the usual no net-flow condition on the external surface of the
sample, we find for the
parameter $T$, Eq.~\rf{Y34}, \beq{tv} T=\frac{\eta
S\phi^2K_0g^2}{k}l_f^2=\frac{l_f^2}{D_T}\,, \eeq where the
coefficient $D_T$ has the dimensions of a diffusion constant and
is given by \beq{dt} D_T\stackrel{\rm def}{=}\frac{k}{\eta
S\phi^2K_0g^2} \:\:\:.\eeq The Poisson size of the fluid patch,
$l_f$, is defined by \beq{lf} l_f^2\stackrel{\rm
def}{=}\frac{1}{V_f}\int_{V_f}dV\tilde{\Phi}\,. \eeq The sole
effect of the membrane stiffness, $W$, in this limit, is thus to
renormalize the coefficient $D_T$ from that which held when $W
= 0$. We have verified numerically that the results derived
in this section are identical to those presented in Sec.~\ref{fa}
when one of the fluids is taken to be a vacuum.

\section{Examples}
\label{ex}

It is straightforward to repeat the calculations of Ref.~\cite{johnson01}
for $\tilde{K}(\omega)$ in which the patchy geometries are
either that of periodic slabs or concentric spheres. In the slab
geometry, region 1 is a layer of thickness $L_1$ and region 2 is a
layer of thickness $L_2$, periodically repeated. Here, $S_1 =
L_1/(L_1+L_2)$ and $A/V = 2/(L_1+L_2)$. In the concentric-spheres
geometry, region 1 is a sphere of radius $R_a$ surrounded by region
2 of outer radius $R_b$: $S_1 = (R_a/R_b)^3$ and $A/V =
3R_a^2/R_b^3$.

Equations (B.4) and (B.8) in Ref.~\cite{johnson01} are still correct as
written for the slab and the sphere geometries, respectively.
However, the pore pressure is no longer continuous across the
patch boundary but it obeys Eq.~\rf{dpf} instead, leading to a
slightly different matrix equation to be solved. The results of
such numerical calculations are presented in Fig.~\ref{fig:Y3},
using the parameter set of \cite{table} at a water saturation $S_w =
0.9$.
The horizontal dashed lines in the left column are $K_{\rm BGH}$, $K_0$,
and $K_{\rm BGW}$ from high to low, respectively. At low frequencies,
we see that the numerically exact results do indeed approach
$K_0$, as per Eq.~\rf{Y30}. $K_{\rm BGW}$ has the same value in all
three geometries, because the saturation values were chosen to be
the same. Similarly for $K_{\rm BGH}$. The values for $K_0$, however,
are different in the three cases because the surface to volume
ratios for the different patch geometries are different. The
parameter $\tilde{W}$ in Eq.~\rf{Y19} has a different
value in each case and so, therefore, do the values of $K_0$.

As we mentioned earlier, the low-frequency parameter $T$ is
determined by a differential equation \rf{Y33}, which is, in
form, identical to that which existed when $W=0$ \cite{johnson01}.
This means that the previously derived results for the slab and the sphere
geometries are virtually unchanged. For the sphere, we have from
Eq.~(40) of Ref.~\cite{johnson01}
\beqa{B16a}
T & = & \frac{K_{0} \phi^2}{30 k R_b^3} \{[3 \eta_2
g_2^2+5(\eta_1-\eta_2)g_1g_2
-3\eta_1g_1^2]R_a^5 \nonumber \\
&&-15\eta_2 g_2 (g_2-g_1)R_a^3 R_b^2 +5g_2 [3\eta_2
g_2-(2\eta_2+\eta_1)g_1] R_a^2R_b^3 -3\eta_2g_2^2R_b^5\}\,.
\eeqa
Similarly, the expression appropriate to the slab geometry
is Eq.~(41) of Ref.~\cite{johnson01}:
\beq{B16b}
T = -\frac{K_{0}\phi^2}{6k(L_1+L_2)}\{\eta_1g_1^2L_1^3 +
3\eta_1g_1g_2L_1^2L_2 + 3\eta_2g_1g_2 L_1L_2^2 +\eta_2g_2^2
L_2^3\}\,.
\eeq
Of course, the correct value for the static
modulus, $K_0$, Eq.~\rf{Y22}, and the correct values for $g_j$,
Eq.~\rf{Y35}, must be used. Subject to this clarification, the
low-frequency attenuation implied by the second term in Eq.~\rf{Y30}
is shown as a dotted line; we see that the numerically computed
results asymptote correctly in this limit.
It was noted in Sec.~\ref{fa} that the high-frequency asymptote for
$\tilde{K}(\omega)$ is unaffected by a nonzero value for $W$. This
limit is also plotted as dotted lines in Fig.~\ref{fig:Y3}.

\section{Theoretical model for $\tilde{K}(\omega)$}
\label{tm}

A key observation by Johnson \cite{johnson01} was that all singularities
and branch cuts of the bulk modulus $\tilde{K}(\omega)$ must lie
on the negative imaginary axis, when analytic continuation into
the complex plane is performed. This stems from the diffusive
nature of the relaxation mechanism. In Appendix~\ref{a2}, we show
that this conclusion stays valid for a nonzero value for the
membrane stiffness, leading us to suggest that the
simple analytical formula proposed earlier \cite{johnson01} may also
work in the present context. We wish to have a function
which has singularities only on the negative imaginary axis and
which has the limiting properties of Eqs.~\rf{Y30} and \rf{hf}.
That is, we propose the following function should work well as
compared against exact numerical results: \beq{am}
\tilde{K}(\omega)=K_\infty-\frac{K_\infty-K_0}{1-\zeta\left(1-\sqrt{1-i\omega\tau/\zeta^2}\right)}\,,
\eeq where consistency with Eqs.~\rf{Y30} and \rf{hf} requires
\beq{B18} \tau=\left[\frac{K_\infty-K_0}{K_\infty G}\right]^2 \eeq
and \beq{B19}
\zeta=\frac{(K_\infty-K_0)^3}{2K_0K_\infty^2TG^2}=\frac{(K_\infty-K_0)}{2K_0}\frac{\tau}{T}\,.
\eeq In Eq.~\rf{am} the branch cut in the definition of the square
root function, $\sqrt{Z}$, is taken to be along the negative real
$z$ axis.

For the specific cases of the slab and concentric spheres
geometries, the parameter $T$ is computed directly from
Eqs.~\rf{B16a}, \rf{B16b} as the case may be. $G$ is evaluated using
Eq.~\rf{g}. The resulting predictions from Eq.~\rf{am} are shown as
dashed curves in Fig.~\rf{fig:Y3}, where one can see that it does an excellent
job of interpolating throughout the entire frequency range.

\section{Conclusions}
\label{con}

In summary, we have extended the theory of Johnson \cite{johnson01} to
consider effects of the capillary forces on the acoustics of
porous media saturated with two different fluids. Using a
well-known expression for the pore-pressure discontinuity across
the patch boundary, Eq.~\rf{dpf}, we have shown that the finite
membrane stiffness, $W$, does not affect the analytic structure of
the frequency-dependent bulk modulus $\tilde{K}(\omega)$. In
addition, the high-frequency asymptote of $\tilde{K}(\omega)$
does not depend on $W$, as long as $W$ is sufficiently small [see
Eqs.~\rf{co}, \rf{ow}], so that the high-frequency limit of our
theory can be reached at all. The effect of the capillary forces
is thus to rescale the static bulk modulus, $K_0$, and the
low-frequency correction parameter, $T$, Eqs.~\rf{Y22} and
\rf{Y34}, respectively. Furthermore, there are no new geometrical
parameters introduced by the membrane stiffness: For example, when
one of the fluids is a gas, the two relevant topological
parameters are the sample-volume to patch-surface ratio and the
Poisson size of the fluid patch, $l_f$, Eq.~\rf{lf}. In the limit
$W\rightarrow0$, results of Ref.~\cite{johnson01} are recovered. In
the limit $W\rightarrow\infty$, $\tilde{K}(\omega)\equiv K_{\rm
BGH}$, Eq.~\rf{bgh}, for all frequencies.

{\acknowledgments

This work was supported in part by the NSF Grant DMR 02-33773.}

\appendix

\section{Capillary correction in the high-frequency limit}
\label{a1}

In this Appendix we show that the pore pressure discontinuity
$\delta p_f$ across the patch boundary, see Eq.~\rf{sw}, scales as
$\omega^{-1/2}$ at high frequencies, $\omega \rightarrow \infty$,
so that the membrane stiffness does not modify the high-frequency
limit. The pressure equilibration, Eq.~\rf{sw}, is provided
by the slow longitudinal waves on both sides of the patch
boundary. In order to find equations of motion for the slow wave
inside either patch, we take divergence of Eqs.~\rf{m1}, \rf{m2}
and use relations \rf{bio}, \rf{bio2}: \beqa{eqm}
\nabla^2\left[(P_k+Q_k)\epsilon^{(k)}_{ll}+(R_k+Q_k)E^{(k)}_{ll}\right]&=&0\,,\\
\nabla^2\left[Q_k\epsilon^{(k)}_{ll}+R_kE^{(k)}_{ll}\right]&=&\frac{i\omega\phi^2\eta_k}{k}\left[\epsilon^{(k)}_{ll}-E^{(k)}_{ll}\right]\,.\label{eqm2}
\eeqa
Looking for solutions of the form
\beq{hm}
\begin{array}{lcr}
\epsilon^{(k)}_{ll}&=&c^{(k)}_1e^{-i(q_kx+\omega t)}\\
E^{(k)}_{ll}&=&c^{(k)}_2e^{-i(q_kx+\omega t)}
\end{array}\,,
\eeq we obtain from Eq.~\rf{eqm} the ratio of the amplitudes of
the solid and fluid phases \beq{c12}
\frac{c^{(k)}_1}{c^{(k)}_2}=-\frac{R_k+Q_k}{P_k+Q_k} \eeq and then
from Eq.~\rf{eqm2} the diffusion coefficient \beq{d}
D_k\stackrel{\rm
def}{=}\frac{i\omega}{q_k^2}=\frac{k}{\phi^2\eta_k}\frac{P_kR_k-Q_k^2}{P_k+2Q_k+R_k}\,.
\eeq The above equations are valid on both sides of the patch
boundary. The coefficients of pressure correction $A_k$,
Eqs.~\rf{sw}, at the interface are then given by \beq{Ak}
A_k=-\frac{1}{\phi}\left[Q_kc^{(k)}_1+R_kc^{(k)}_2\right]\, \eeq
where we used Eq.~\rf{bio2} for the additional pressure exerted by
the slow waves. The pressure drop $\delta p_f$ across the
interface is related to the relative fluid displacement through
Eq.~\rf{dpf}, and using Eq.~\rf{m2} we thus obtain \beq{pfn}
\delta
p_f=\left(p_{f1}+A_1\right)-\left(p_{f2}+A_2\right)=-\frac{q_1kW}{\eta_1\omega}A_1\,.
\eeq Finally, we require the fluid-flux continuity across the
patch boundary: \beq{fc}
\frac{q_1}{\eta_1}A_1=-\frac{q_2}{\eta_2}A_2\,. \eeq
Eqs.~\rf{c12}, (\ref{Ak})-(\ref{fc}) now constitute a linear system
of 6 equations with six unknowns: $c^{(k)}_1$, $c^{(k)}_2$, and
$A_k$. The resulting expression for $\delta p_f$, Eq.~\rf{pfn},
simplifies considerably if one patch is filled with a vacuum. In
this case, the pressure discontinuity at the fluid boundary,
relative to the difference in pore pressure relatively far from
the interface is given by \beq{pfvn} \frac{\delta p_f}{\Delta
p_f}=\frac{1}{1+\eta\omega/Wkq}=\frac{1}{1+\sqrt{\omega/i\omega_W}}\,,
\eeq where the crossover frequency \beq{co} \omega_W\stackrel{\rm
def}{=}\frac{W^2k^2}{\eta^2D} \eeq is defined in terms of
quantities corresponding to the fluid patch. When
$\omega\gg\omega_W$, $\delta p_f\propto\omega^{-1/2}$ and the
pressure drop across the interface, $\delta p_f$, is negligible in
comparison with the pressure difference inside the patches,
$\Delta p_f$, Eq.~\rf{dp}.

For consistency of our theory, we require that this limit can be
reached while we are still in the low-frequency regime of the Biot
theory and the propagatory modes have wave lengths longer than the
typical patch dimensions, i.e., we need \beq{ow}
\omega_W\ll(\omega_B,\omega_x)\,. \eeq Eqs.~\rf{co} and \rf{ow}
imply that our high-frequency limit is valid only for sufficiently
small membrane-stiffness parameter $W$. With the parameter set
used to generate figure~\ref{fig:Y3} we have $\omega_W/(2\pi) = $
4.5 Hz. However, our theory makes no assumption about the value of
$\omega_W$ as compared against the diffusive crossover frequency
and, indeed, the theory works quite well even if $W$ is so large
that $\omega_W$ is greater than the crossover implied by the
intersection of the high and low frequency asymptotes.

\section{Analytic structure of $\tilde{K}(\omega)$}
\label{a2}

Here we generalize Appendix~C of Ref.~\cite{johnson01} to take into
account the additional term due to the finite membrane stiffness
$W$ in Eq.~\rf{m2g}. We want to investigate analytic properties of
the real-valued causal response function $\hat{K}(t)$ in the
frequency domain: \beq{JC1} \tilde{K}(\omega)\stackrel{\rm
def}{=}\int_0^\infty\hat{K}(t)e^{i\omega t}dt\,. \eeq It is clear
that $\tilde{K}(\omega)$ is analytic everywhere in the upper-half
complex $\omega$ plane, and we will further show that all the
zeros, singularities, and branch cuts lie on the negative
imaginary axis.

In the following we consider an eigenvalue problem in which all quantities
vary as $\mathbf{u}(\mathbf{r})e^{-i\omega_n t}$. Assume
\beq{ts} \int dS\mathbf{
u}\cdot\mbox{\boldmath$\tau^\ast$}\cdot\mathbf{\hat{n}}=0\,, \eeq
as would be in the case of a zero, singularity, or branch cut of
$\tilde{K}(\omega)$ \cite{johnson01}. We can then show using
Eq.~\rf{m1} and the Biot relations between stresses and strains,
Eqs.~\rf{bio} and \rf{bio2}, that $\int dV p_f\tau^\ast_{ll}$ is
real-valued. This is true since all these equations are unaffected
by the capillary forces, and we thus reproduce Eq.~(C5) of
Ref.~\cite{johnson01}: \beq{JC5} \int dVp_f\tau^\ast_{ll}=\int
dV\frac{1}{K_b/K_s-1}\left[\frac{1}{3}|\tau_{ll}|^2+\frac{3K_b}{2N}\left(\mathbf{
DD^\ast}\right)_{ll}\right]\,, \eeq where
$D_{ij}=\tau_{ij}-(1/3)\tau_{ll}\delta_{ij}$ is the deviatoric
part of the stress tensor. Finally, multiplying Eq.~\rf{m2g} by
$\mathbf{U^\ast}-\mathbf{u^\ast}$ and integrating it over the
entire volume of the sample, we obtain \beq{in} i\omega_n\int
dV\frac{\eta\phi}{k}|\mathbf{U}-\mathbf{u}|^2=\int
dVp_f(\epsilon_{ll}^\ast-E_{ll}^\ast)+W\phi\int
dV|\mathbf{\hat{n}}\cdot(\mathbf{U}-\mathbf{u})|^2\delta(R)\,.
\eeq As the volumetric strains, $\epsilon_{ll}$ and $E_{ll}$, are
given by linear combinations of $p_f$ and $\tau_{ll}$ [via
inverting Eqs.~\rf{bio}, \rf{bio2}], the right-hand side of
Eq.~\rf{in} is real-valued and, therefore, $\omega_n$ is pure
imaginary. This completes the proof.

\clearpage

\begin{figure}[h!]
\caption{Dependence of the static modulus on
$\tilde{W}$ for a sample saturated with water, $S_w =
0.7$ and air, $S_a = 0.3$.  Parameters are listed in \cite{table}.  The
two results for the static modulus, Eqs.~\rf{Y22} and \rf{K0},
are seen to be nearly identical in this case in which one of the fluids
is gas. $K_{\text{BGH}}>K_{\text{BGW}}$.} \label{fig:Y1}
\end{figure}

\begin{figure}[h!]
\caption{The static moduli as a function of the water
saturation, $S_w$. Parameter set as in Fig.~\ref{fig:Y1}.
$\tilde{W}=1$~GPa is held constant.
Again, $K_0$ is calculated in two different
ways. $K_{\text{BGH}}\geq K_{\text{BGW}}$.} \label{fig:Y2}
\end{figure}

\begin{figure}[h!]
\caption{Dispersion (${\rm Real}[\tilde{K}(\omega)]$)
and attenuation
[$1/Q_K=-{\rm Imag}\tilde{K}(\omega)/{\rm Real}\tilde{K}(\omega)$]
due to the patchy-saturation effect for an
air/water combination, $S_a=10$\%, in three different situations.
Solid curves are numerically exact solutions, dotted curves are the
high- and low-frequency limiting expressions, and dashed curves
represent the analytic expression, Eq.~(\ref{am}). Horizontal dashed lines
are (in order of decreasing value) $K_{\text{BGH}}$,
$K_0(W=30~{\rm GPa/m})$, and $K_0(W=0)$.
(A) Top Row: The gas is
located in an inner sphere, $R_a=4.642$~cm, surrounded by a
shell of water, $R_b=10$~cm. (B) Middle Row: The roles are
reversed, $R_a=9.655$~cm, $R_b=10$~cm. (C) Bottom Row: A
periodic slab geometry, $L_a=2$~cm, $L_w=18$~cm. The other
material parameters are listed in \cite{table}. In all cases, the
membrane stiffness is $W=30$~GPa/m. The values of $\zeta$, Eq.~\rf{B19},
are shown.} \label{fig:Y3}
\end{figure}

\newpage

\includegraphics[width=\textwidth]{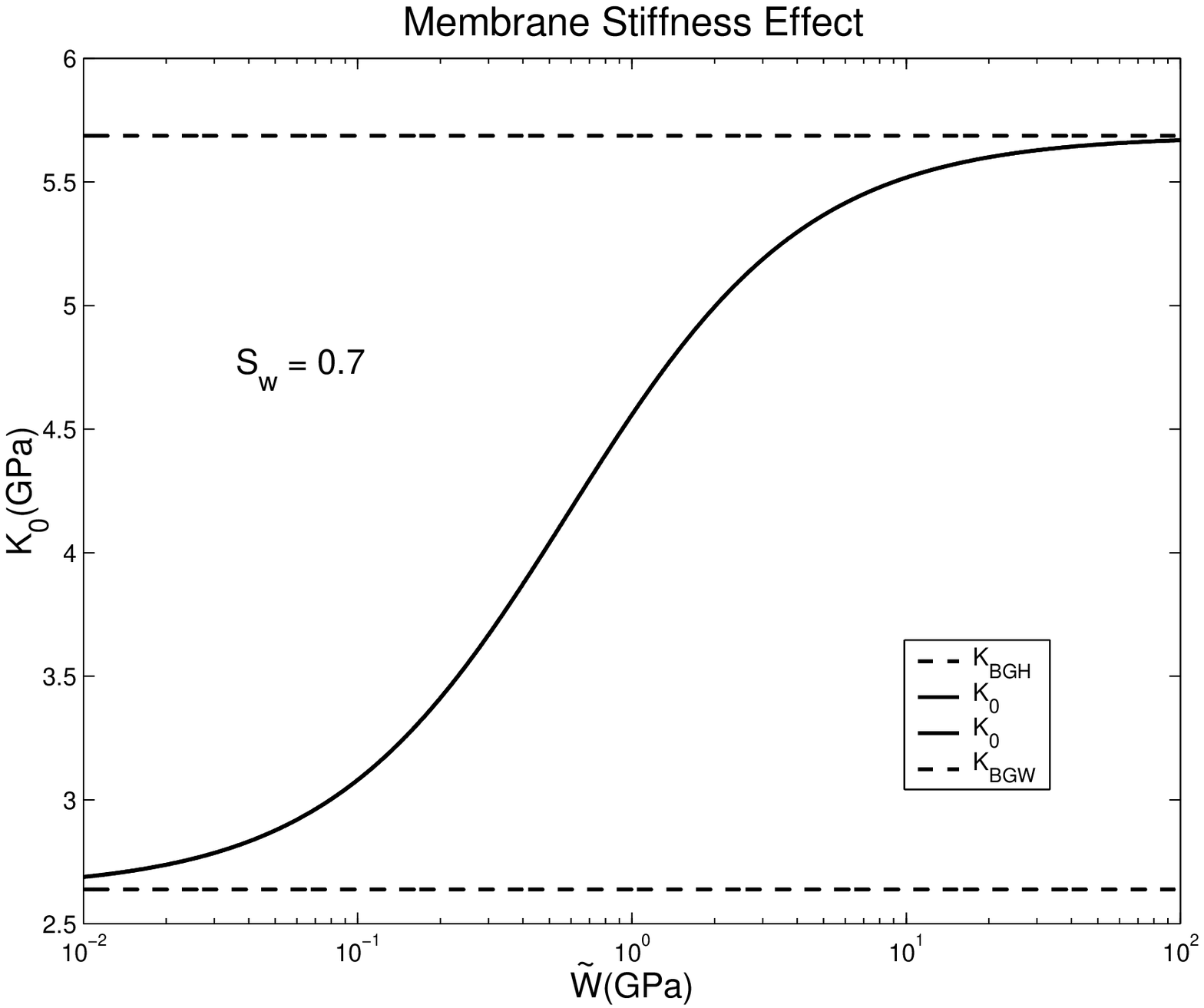}
\vspace{2cm}

{\Large Figure 1. Y. Tserkovnyak and D. L. Johnson, J. Acoust. Soc. Am.}

\newpage

\includegraphics[width=\textwidth]{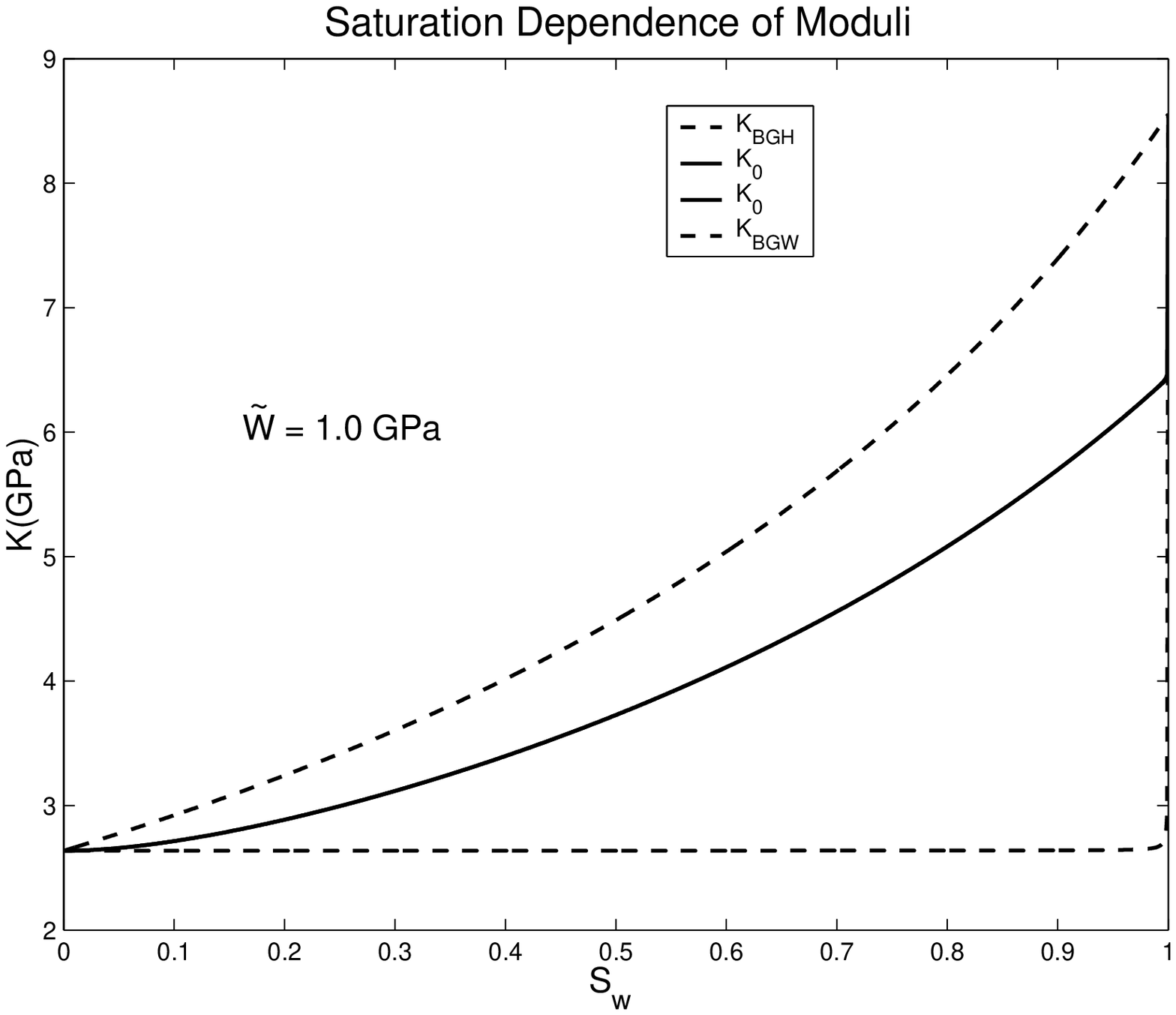}
\vspace{2cm}

{\Large Figure 2. Y. Tserkovnyak and D. L. Johnson, J. Acoust. Soc. Am.}

\newpage

\includegraphics[width=\textwidth]{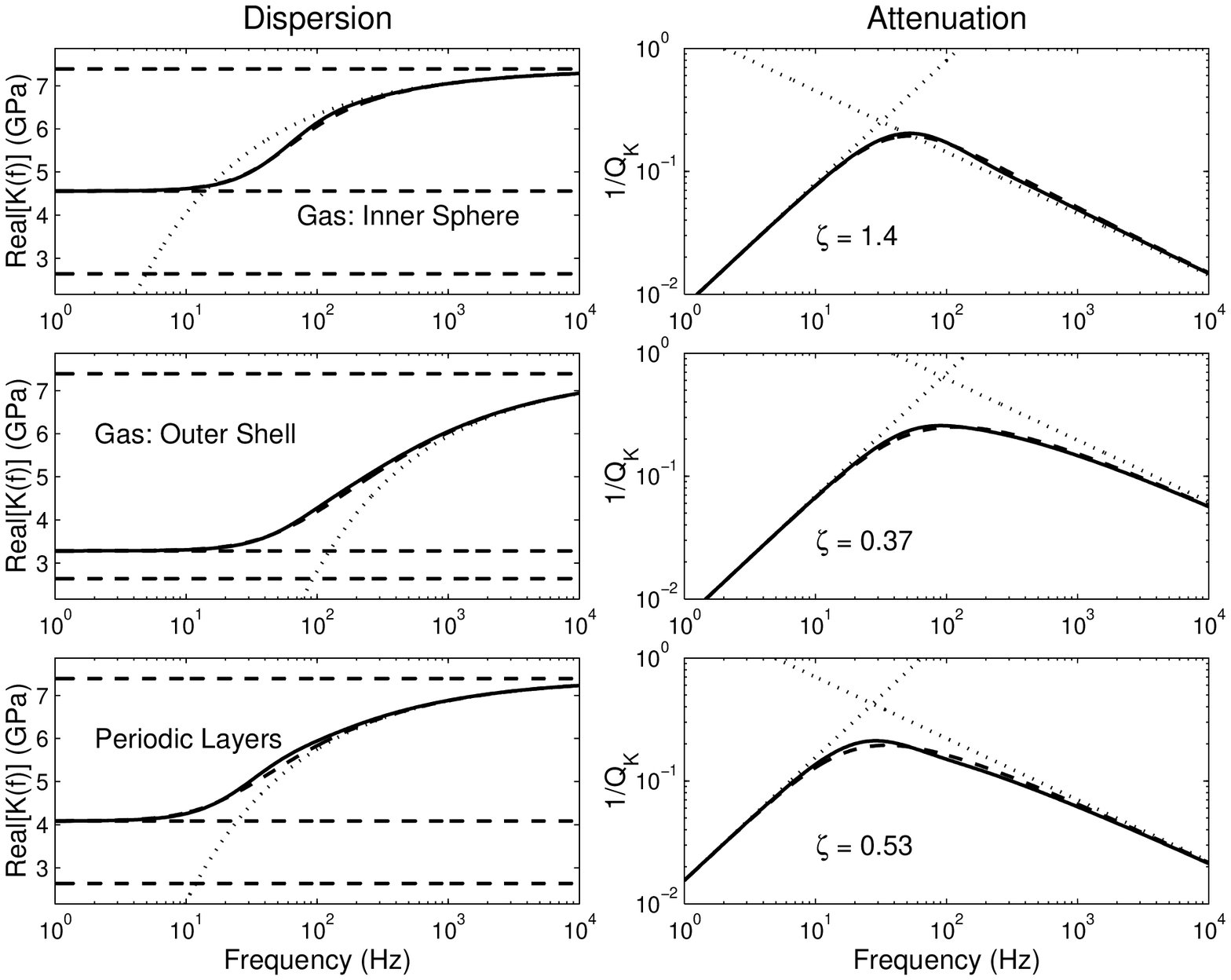}
\vspace{2cm}

{\Large Figure 3. Y. Tserkovnyak and D. L. Johnson, J. Acoust. Soc. Am.}


\begin{thebibliography}{}
\bibitem{table} Values of input parameters for the
calculation of patchy-saturation effects (from Ref.~\cite{johnson01}):
$\phi=0.284$, $K_s=35$~GPa, $K_b=2.637$~GPa, $N=1.740$~GPa, $k=10^{-13}$~m$^2$, $K_f~\text{(water)}=2.25$~GPa, $\eta~\text{(water)}=10^{-3}$~Pa s, $K_f~\text{(air)}=10^5$~Pa, $\eta~\text{(air)}=10^{-5}$~Pa s, $\sigma$(water-air)$=7.3\times10^{-2}$~N/m. From Eq.~(\ref{YT4}) we can then find $W=30$~GPa/m using $s=0.04$ and from Eq.~(\ref{Y19b}) $\tilde{W}=1$~GPa using $V/A=0.4$~m.
\end{thebibliography}
\end{document}